\documentstyle[epsf,twocolumn]{article}
\begin{document}

\begin{titlepage}
\begin{center}
{\Huge
Decay Constant of Pseudoscalar Meson in the Heavy Mass Limit}
$$
$$
$$
$$

{\Large
V.V. Andreev\\
Gomel State University, Physics Department,
Gomel, 246699,Belarus.  \\
E-mail: andreev@gsu.unibel.by\\}
$$
$$
{\large
Published in Proceedings
7 Annual Seminar "Nonlinear Phenomena In Complex System" (NPCS'98)\\
( Minsk, Belarus, 1998  )\\ .}
\end{center}
$$
$$
$$
$$
\begin{center}
{\Huge Abstract}
\end{center}

\begin{center}

{\Large
The leptonic decay constant of the pseudoscalar mesons a
calculated by use of the
relativistic constituent quark model constructed on the point form
of Poincare-covariant quantum mechanics.
We discuss the role relativistic corrections for decay constants of
pseudoscalar mesons with heavy quarks. We consider the heavy mass
limit of decay constant for two-particle system with equal masses.}
\end{center}
\end{titlepage}

\newpage
{\Large Decay Constant of Pseudoscalar Meson
in the Heavy Mass Limit\\ }
$$
$$
{\large V.Andreev\\
Gomel State University, Physics Department,\\
Gomel, 246699,Belarus.  \\
E-mail: andreev@gsu.unibel.by\\ }

\section{\bf Introduction}

According to the structure of the current interaction, electroweak decays of
hadrons can be divided into following classes: leptonic decays,in which the
quarks of the decaying hadron annihilate each other and only leptons appear
in the final state; semileptonic decays, in which both leptons and hadrons
appear in the final state; photons decays,in which the final state consists
of photons only; radiative transitions between hadrons, in which hadrons and
photon are caused by hadron decays. non-leptonic decays,in which the final
state consists of hadrons only. Over the last decade, a lot of information
on hadron decays has been collected in experiments at $e^{+}e^{-}$ and
hadron colliders. This has led to a rather detailed knowledge of the flavour
sector of the Standard Model and many of the parameters associated with it.
In this work we investigate the decay constant of the mesons with spinor
quarks in the heavy mass limit.

\section{$q\overline{q}$ bound state in the point form of RQM}

There are three forms of the dynamics in the relativistic quantum mechanics
(RQM), called instant ,point,light-front forms \cite{Di1}. In this work we
use point form of the RQM \cite{Gross}. The description in the point form
implies that the generators of the Poincare group ${\hat M}^{\mu\nu }$ are
the same as for noninteracting particles and bound systems. Interaction
terms can be present only in the four-momentum operators ${\hat P}_\mu $,
but the four-velocities of bound and free-particle systems are equal.

The momenta $\vec p_1$, $\vec p_2$ of the quarks with the masses $m_1$ and $
m_2$ of relativistic system can be transformed to the total $\vec P$ and
relative momenta $\vec k$ to facilitate the separation of the center mass
motion:
\[
\vec P_{12} =\vec p_1+\vec p_2,
\]
\begin{equation}
\overrightarrow{k} =\overrightarrow{p_1}+\frac{\overrightarrow{P_{12}}}{M_0 }
\left( \frac{\left( \overrightarrow{P_{12}}\overrightarrow{p_1}\right) }{
\omega _{M_0}\left( \overrightarrow{P_{12}}\right) +M_0}+\omega _{m_1}\left(
\overrightarrow{p_1}\right) \right) ,  \label{veck}
\end{equation}
where $M_0=\omega _{m_1}\left( \overrightarrow{k}\right) +\omega
_{m_2}\left( \overrightarrow{k}\right) ,\omega _{m_1}\left( \overrightarrow{
p_1}\right) =\sqrt{\overrightarrow{p_1}^2+m_1^2}$.

The solution of the eigenvalue problem will lead to eigenfunction of the
form
\[
_0\left\langle \vec V_{12}\mu ,\left[ J\hskip 2ptk\right] ,(l s)\right.
\left| \vec V\mu ,\left[ J\hskip 2ptM\right] \right\rangle =
\]
\begin{equation}
=\delta_{J J^{\prime }} \delta _{\mu \mu^{\prime }} \delta (\vec V-\vec
V_{12}) \Psi ^{J\mu }\left( k\hskip 2pt l\hskip 2pt s\hskip 2pt; M \right)
\end{equation}
with the velocities of bound system $\vec V={\vec P}/{M}$ and noninteracting
system $\vec V_{12}={\vec P_{12}}/{M_0}$. The function $\Psi ^{J\mu }\left(
k \hskip 2ptl\hskip 2pts\hskip 2pt;M\right) $ satisfies in the point form a
following equation \cite{Pol1}:
\[
\sum_{l^{\prime }s^{\prime }}\int\limits_0^\infty <k\hskip 2ptl\hskip
2pts\parallel W^J\parallel k^{\prime }\hskip 2ptl^{\prime }\hskip
2pts^{\prime }>\Psi ^J(k^{\prime }\hskip 2ptl^{\prime }\hskip 2pts^{\prime
}; \hskip 2ptM)k^{\prime 2}dk^{\prime }+
\]
\begin{equation}
+k^2\Psi ^J(k\hskip 2ptl\hskip 2pts;M)=\eta \Psi ^J(k\hskip 2ptl\hskip 2pt
s;M)  \label{maineq}
\end{equation}
with reduced matrix element of operator $\hat W$.

In the point form the meson state is defined by as state of on-shell quark
and antiquark with the wave function $\Psi ^{J\mu }\left( k\hskip 2pt l
\hskip 2pt s\hskip 2pt;M\right) $
\[
\left| \overrightarrow{P}\mu \hskip 2pt\left[ JM\right] \right\rangle =\sqrt{
\frac M{\omega _M\left( \overrightarrow{P}\right) }}*
\]
\[
\ast \sum_{ls\lambda _1\lambda _2}\int d^3k\sqrt{\frac{\omega _{m_1}\left(
\overrightarrow{p_1}\right) \omega _{m_2}\left( \overrightarrow{p_2}\right)
}{\omega _{m_1}\left( \overrightarrow{k}\right) \omega _{m_2}\left(
\overrightarrow{k}\right) }}\Psi ^{J\mu }\left( kls;M\right)
\]
\[
\sum_{m\lambda }\sum_{\nu _1\nu _2}\left\langle s_1\nu _1,s_2\nu _2\right|
\left. s\lambda \right\rangle \left\langle lm,s\lambda \right| \left. J\mu
\right\rangle Y_{lm}\left( \theta ,\phi \right)
\]
\begin{equation}
D_{\lambda _1\nu _1}^{1/2}\left( \overrightarrow{n}\left( p_1,P\right)
\right) D_{\lambda _2\nu _2}^{1/2}\left( \overrightarrow{n}\left(
p_2,P\right) \right) \left| p_1\lambda _1\right\rangle \left| p_2\lambda
_2\right\rangle  \label{state}
\end{equation}
where $\left\langle s_1\nu _1,s_2\nu _2\right| \left. s\lambda \right\rangle
$, $\left\langle lm,s\lambda \right| \left. J\mu \right\rangle $ are
Clebsh-Gordan coefficients of $SU(2)$-group, $Y_{lm}(\theta ,\phi )$ -
spherical harmonic with spherical angle of $\vec k$. Also, in Eq.(\ref{state}
) $D^{1/2}\left( \overrightarrow{n}\right) =1-i\left( \overrightarrow{n}
\overrightarrow{\sigma }\right) /\sqrt{1+\overrightarrow{n}^2}$is $D$
-function of Wigner rotation, which determined by vector-parameter $
\overrightarrow{n(}p_1,p_2)=\overrightarrow{u_1}\times \overrightarrow{u_2}
/(1-\left( \overrightarrow{u_1}\overrightarrow{u_2}\right) )$ with $
\overrightarrow{u}=\overrightarrow{p}/\left( \omega _m\left( \overrightarrow{
p}\right) +m\right) $.

\section{\bf Leptonic decay constant }

The leptonic decay constant for pseudoscalar meson is defined by
\begin{equation}
\left\langle 0\left| \hat J^\mu \left( 0\right) \right| \overrightarrow{P}
,M\right\rangle =i\left( 1/2\pi \right) ^{3/2}\frac 1{\sqrt{2\omega _M\left(
\overrightarrow{P}\right) }}P^\mu f_p,  \label{deconst}
\end{equation}
where $\hat J^\mu (0)$ is the operator axial-vector part of the charged weak
current. Using Eq.(\ref{state}) and Eq.(\ref{deconst}) we found in the point
form dynamics, that \cite{And1}
\[
f_p=\frac{N_c}{\pi \sqrt{2}}\int_0^\infty dk\hskip 2ptk^2\sqrt{\frac{
M_0^2-(m_1-m_2)^2}{\omega _{m_1}\left( \overrightarrow{k}\right) \omega
_{m_2}\left( \overrightarrow{k}\right) }}*
\]
\begin{equation}
\ast \frac{\left( m_1+m_2\right) }{M_0^{3/2}}\Psi \left( k,M\right) ,
\label{dec1}
\end{equation}
where $N_c$-number of colors, $m_1$ and $m_2$ are the respective masses of
the two quarks. The wave function for pseudoscalar meson have the
normalization
\[
\int_0^\infty dk\hskip 2ptk^2\hskip 2ptN_c\left| \Psi \left( k,M\right)
\right| ^2=1.
\]
When $m_1=m_2=m_Q$, the leptonic decay constant is defined by

\begin{equation}
f_p=\frac{2N_cm_Q}\pi \int_0^\infty \frac{dk\hskip 2ptk^2\Psi \left(
k,M\right) }{\omega _{m_Q}^{3/2}\left( \overrightarrow{k}\right) }.
\label{dec1d}
\end{equation}

The equation for the bound $q\bar q$ states (\ref{maineq}) in the RQM is
relativistic equation with effective potential $W$ . However, it is hard
problem to obtain wave function $\Psi (k,M)$ as solution of this equation.
Therefore, we use simple model wave function depending on length scale
parameter $1/\beta $:

\begin{equation}
\Psi (k\hskip 2pt,M)\equiv \Psi (k\hskip 2pt,\beta )=2/(\sqrt{N_c}\beta
^{3/2}\pi ^{1/4})exp(-\frac{k^2}{2\beta ^2}) .  \label{wv1}
\end{equation}

Using the equations (\ref{dec1d}) and (\ref{wv1}), one can see that

\begin{eqnarray}
f_\pi &=&\frac{\sqrt{N_c}\beta }{\pi ^{5/4}\Gamma \left( -\frac 14\right) W}
\nonumber \\
&&(2^{3/4}\Gamma \left( -\frac 14\right) \Gamma \left( \frac 34\right) {
_1F_1 }\left( \frac 34;\frac 14;\frac 1{2W^2}\right) -  \nonumber \\
&&\frac{2\sqrt{\pi }}{W^{3/2}}\Gamma \left( -\frac 34\right) {_1F_1}\left(
\frac 32;\frac 74;\frac 1{2W^2}\right) ),  \label{dec1a}
\end{eqnarray}
with $W=\beta /m_Q$, hypergeometric function $_1F_1(a;b;z)$ and $\Gamma (z)$
-Gamma function.

We now consider the heavy mass limit of (\ref{dec1}). This limit is defined
as $m_1$, $m_2\longrightarrow \infty $ with $V=P/M$ fixed. The starting
point in the construction of the effective theory with the heavy quarks
(HQET) is the observation that a heavy quark bound inside a hadron moves
more or less with the hadron's velocity $V$. Its momentum can be written as
\begin{equation}
p_Q=m_QV+\widetilde{k},  \label{hq1}
\end{equation}
where the components of the so-called residual momentum $\widetilde{k}$ are
much smaller than $m_Q$. Interactions of the heavy quark with light degrees
of freedom change the residual momentum by an amount of order $\widetilde{k}
\sim \Lambda _{QCD}\simeq 1/R_{Hadron}$, but the corresponding changes in
the heavy-quark velocity vanish as $\Lambda _{QCD}/m_Q\longrightarrow 0$. In
the system of the center mass we are obtained, that relative momentum $
\overrightarrow{k}$ (\ref{veck}) and residual momentum$\overrightarrow{
\widetilde{k}}$ are equal and therefore, the heavy mass limit in the point
form is given by
\begin{equation}
\left| \overrightarrow{k}\right| \leq \Lambda _{QCD}\ll m_Q.  \label{hlim}
\end{equation}
The nonrelativistic variant furnishes the following relationship for
leptonic decay constant (\ref{dec1}):

\begin{eqnarray}
f_{nonrel} &=&\frac{2N_cm_Q}\pi \int_0^{\Lambda _{QCD}}dk \hskip 2pt k^2
\nonumber \\
&&(1-\frac{3 k^2}{4 m_Q^2}+\frac{21 k^4}{32 m_Q^4})\Psi _{nonrel}\left(
k,M\right) ,  \label{dec2}
\end{eqnarray}
where $\Psi _{nonrel}\left( k,M\right) $ have the normalization
\[
\int_0^{\Lambda _{QCD}}dk\hskip 2pt k^2\hskip 2pt N_c\left| \Psi
_{nonrel}\left( k,M\right) \right| ^2=1.
\]
The parameter $\beta $ can be estimate from mean square radius (MSR) $
\left\langle r^2\right\rangle _{nonrel}=1/\Lambda _{QCD}^2$ of the meson
with the heavy quarks. In the nonrelativistic approximation the MSR is $
\left\langle r^2\right\rangle _{nonrel}=3/8\beta ^2$ and we obtain the
relationship between $\beta $ and $\Lambda _{QCD}$:

\begin{equation}
\Lambda _{QCD}=\sqrt{\frac 83}\beta .  \label{lb}
\end{equation}

The nonrelativistic wave function can be choose the form:

\begin{equation}
\Psi _{nonrel}\left( k,M\right) \sim \Psi \left( k,M\right) \sim exp(-\frac{
k^2}{2\beta ^2}),  \label{nonfun}
\end{equation}
since the model wave function (\ref{wv1}) has not the small parameter $
\Lambda _{QCD}/m_Q$ (or $W=\beta /m_Q$).Using (\ref{dec2}),(\ref{lb}) and (
\ref{nonfun}) we obtain the following result for $f_{nonrel}$:

\begin{equation}
f_{nonrel}\approx \sqrt{N_c}\beta \sqrt{W}(0.72 -0.73 W^2+1.09 W^4).
\label{dec1c}
\end{equation}

Let us discuss in brief the role of relativistic corrections in leptonic
decay constant of pseudoscalar meson with heavy quarks. This effect can be
extracted easily. Using asymptotic limit for Kummer's function ($
1/W\longrightarrow \infty $)we found, that decay constant (\ref{dec1a}) can
be written as

\begin{eqnarray}
f_p &\approx &\frac{\sqrt{N_c}\beta \sqrt{W}}{\pi ^{3/4}16\sqrt{2}}\left(
32-72 W^2+315 W^4\right)  \nonumber \\
&\approx &\sqrt{N_c}\beta \sqrt{W}(0.60 -1.35 W^2+5.90 W^4).  \label{dec1b}
\end{eqnarray}

Comparison of series (\ref{dec1b}) and (\ref{dec1c}), that the factors at
addenda of these series differ, especially second and third term of a
series. Just, these addenda also give corrections to the effective theory of
heavy quarks. If we let's assume, that the parameter $\Lambda _{QCD}=a\beta $
with $a=\sqrt{2}$, the first terms of series practically coincides for two
variants
\begin{equation}
f_{nonrel}\approx \sqrt{N_c}\beta \sqrt{W}(0.60-0.47 W^2+0.55 W^4),
\label{de1ca}
\end{equation}
but the second and third terms nevertheless essentially differ. Practically
we compare two approaches of an evaluation of relativistic corrections for
the effective theory of heavy quarks: the first approach follows from an
exact solution of a problem with a consequent passage to the limit of heavy
quarks; the second approach is based on an approximate solution of a
problem; In the second approach, and such approach, as we see requires
cutting relative momentum by magnitude $\Lambda _{QCD}$ in a quark model of
a meson, relativistic corrections has the smaller value just because of
cutting. Such divergence can be reduced by introduction of a parameter $\mu $
, which $\gg \Lambda _{QCD}$. However, it can be defined a value only using
exact calculation. Therefore use of exact expressions for observable
magnitudies is represented preferable to us, as, the numerical integration
both approximate relations, and exact expressions has an identical order of
complexity.

\section{\bf ACKNOWLEDGMENTS}

This work was supported by grant {\bf N. F96-326 (17.02.97}) from the
Foundation for Fundamental Research of Republic Belarus.


\begin{thebibliography}{9}
\bibitem{Di1}  P.A.M. Dirac Rev.Mod.Phys., {\bf 21}, p.392 (1949)

\bibitem{Gross}  F.Gross Phys.Rev.,{\bf \ 186}, p.1448 (1969); Phys. Rev. D,
{\bf 10}, p.223 (1974). F.M.Lev ''Forms of relativistic dynamics, current
operators and deep inelastic lepton-nucleon scattering'' hep-ph/9505373.

\bibitem{Coest1}  F. Coester and W.N. Polyzou, Phys.Rev. D, {\bf 26},1348,
(1982).

\bibitem{Pol1}  W.N. Polyzou, Annals of Physics {\bf 193}, p.367 (1989).
B.D.Keister, W.N.Polyzou in Advances in Nuclear Physics, edited J.W.Negele
and E.Vogt (Plenum, New York, 1991).

\bibitem{And1}  V.V.Andreev
Proc.3-d Annual Seminar NPCS'94. Editors
V.I.Kuvshinov \& G.Krylov,Minsk 1995. P.141-145.
\end{thebibliography}
\end{document}